\documentclass[aps,pra,showpacs,twocolumn]{revtex4}%
\usepackage{graphics}
\usepackage{amsmath}
\usepackage{graphicx}
\usepackage{amsfonts}
\usepackage{amssymb}%
\begin{document}
\title{Restoring broken entanglement by separable correlations}

\begin{abstract}
We consider two bosonic Gaussian channels whose thermal noise is strong enough
to break bipartite entanglement. In this scenario, we show how the presence of
separable correlations between the two channels is able to restore the broken
entanglement. This reactivation occurs not only in a scheme of direct
distribution, where a third party (Charlie) broadcasts entangled states to
remote parties (Alice and Bob), but also in a configuration of indirect
distribution which is based on entanglement swapping. In both schemes, the
amount of entanglement remotely activated can be large enough to be distilled
by one-way distillation protocols.

\end{abstract}

\author{Gaetana Spedalieri}
\author{Stefano Pirandola}
\email{stefano.pirandola@york.ac.uk}
\affiliation{Computer Science and York Centre for Quantum Technologies, University of York,
York YO10 5GH, United Kingdom}
\maketitle

Entanglement is a fundamental physical resource in quantum information and
computation. Once two parties, say Alice and Bob, share a suitable amount of
entanglement, they can implement a variety of powerful
protocols~\cite{NielsenBook,Mwilde}. In a scheme of direct distribution, there
is a middle station (Charlie) possessing a bipartite system in an entangled
state; one subsystem is sent to Alice and the other to Bob. Alternatively, in
a scheme of indirect distribution, known as entanglement swapping, the
distribution is mediated by a measurement process. Here Alice and Bob each has
a bipartite system prepared in an entangled state. One subsystem is retained
while the other is sent to Charlie. At his station, Charlie detects the two
incoming subsystems by performing a suitable Bell measurement and communicates
the classical outcome back to Alice and Bob. As a result of this process, the
two subsystems retained by the remote parties are projected onto an entangled state.

\begin{figure}[ptbh]
\vspace{-0.4cm}
\par
\begin{center}
\includegraphics[width=0.62\textwidth] {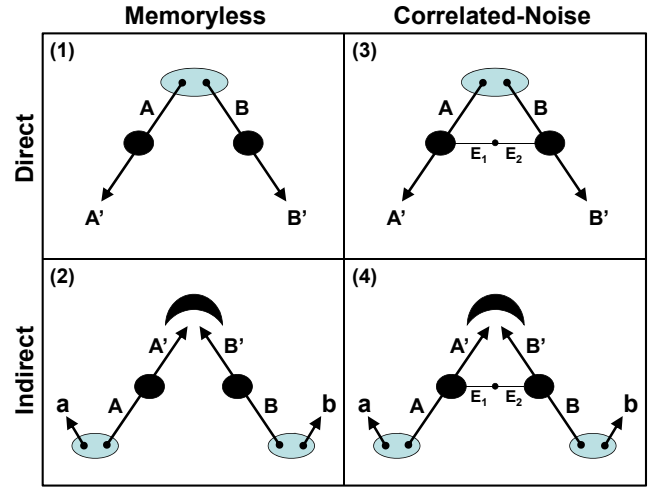}
\end{center}
\par
\vspace{-1.0cm}\caption{Direct and indirect schemes for entanglement
distribution in memoryless and correlated-noise environments. Charlie is the
middle station transmitting to or receiving systems from Alice (left station)
and Bob (right station). Ellipses represent entangled states, black-circles
represent channels, and detectors are Bell measurements. (1) Direct
distribution in a memoryless environment. If the individual channels are EB,
then no distribution of entanglement is possile. We have $A|B^{\prime}$ and
$A^{\prime}|B$ which implies $A^{\prime}|B^{\prime}$. (2) Entanglement
swapping in a memoryless environment. We have that $a|A^{\prime}$ and
$B^{\prime}|b$ implies $a|b$. (3) Direct distribution in a correlated
environment. Despite $A|B^{\prime}$ and $A^{\prime}|B$ we have that
$A^{\prime}-B^{\prime}$ is possible. Surprisingly, this can be realized by a
separable environment. (4) Entanglement swapping in a correlated environment.
Despite $a|A^{\prime}$ and $B^{\prime}|b$ we have that $a-b$ is possible.
Surprisingly, this is realizable by a separable environment.}%
\label{scenario}%
\end{figure}

In both configurations, entanglement distribution is possible as long as the
action of the environment is not too strong. When decoherence is strong enough
to destroy any input entanglement, the environment results into an
entanglement breaking (EB) channel~\cite{EBchannels,HolevoEB}. By definition,
a quantum channel $\mathcal{E}$\ is EB when its local action on one part of a
bipartite state always results into a separable output state. In other words,
given two systems, $A$ and $B$, in an arbitrary bipartite state $\rho_{AB}$,
the output state $\rho_{AB^{\prime}}=(\mathcal{I}_{A}\otimes\mathcal{E}%
_{B})(\rho_{AB})$\ is always separable, where $\mathcal{I}_{A}$ is the
identity channel applied to system $A$ and $\mathcal{E}_{B}$ is the EB channel
applied to system $B$. Thus, if the input systems $A$ and $B$ were initially
entangled (here denoted by the notation $A-B$), the output systems $A$ and
$B^{\prime}$ are separable (here denoted by the notation $A|B^{\prime}$).

The standard model of decoherence is assumed to be Markovian, where the
travelling systems are subject to memoryless channels. For instance, consider
the case of direct distribution depicted in the panel~(1) of
Fig.~\ref{scenario}. In the standard Markovian description, the entangled
state $\rho_{AB}$ of the input systems $A$ and $B$ is subject to a tensor
product of channels $\mathcal{E}_{A}\otimes\mathcal{E}_{B}$. In this case,
there is clearly no way to distribute entanglement if both $\mathcal{E}_{A}$
and $\mathcal{E}_{B}$ are EB channels. Suppose that Charlie tries to share
entanglement with one of the remote parties by sending one of the two systems
while keeping the other (one-system transmission). For instance, Charlie may
keep system $A$ while transmitting system $B$ to Bob. The action of
$\mathcal{I}_{A}\otimes\mathcal{E}_{B}$ destroys the initial entanglement, so
that systems $A$ (kept) and $B^{\prime}$ (transmitted) are separable
($A|B^{\prime}$). Symmetrically, the action of $\mathcal{E}_{A}\otimes
\mathcal{I}_{B}$ destroys the entanglement between system $A^{\prime}$
(transmitted) and system $B$ (kept), i.e., we have $A^{\prime}|B$. Then
suppose that Charlie sends both his systems to Alice and Bob (two-system
transmission). This strategy will also fail since the joint action of the two
EB\ channels is given by the tensor product $\mathcal{E}_{A}\otimes
\mathcal{E}_{B}=(\mathcal{E}_{A}\otimes\mathcal{I}_{B})(\mathcal{I}_{A}%
\otimes\mathcal{E}_{B})$. In other words, since we have one-system EB
($A|B^{\prime}$ and $A^{\prime}|B$) then we must have two-system EB
($A^{\prime}|B^{\prime}$).

The previous reasoning can be extended to the case of indirect distribution as
shown in panel (2) of Fig.~\ref{scenario} involving a Bell measurement by
Charlie. Since the environment is memoryless ($\mathcal{E}_{A}\otimes
\mathcal{E}_{B}$), we have that the absence of entanglement before the Bell
measurement ($a|A^{\prime}$ and $B^{\prime}|b$) is a sufficient condition for
the swapping protocol to fail, i.e., the remote systems $a$ and $b$ remains
separable ($a|b$). Similarly to the previous case, if one-system transmission
does not distribute entanglement, then two-system transmission cannot lead to
entanglement generation via the swapping protocol.

Here we discuss how the previous implications for direct and indirect
distribution of entanglement are false in the presence of a correlated-noise
environment: Two-system transmission can successfully distribute entanglement
despite one-system transmission being subject to EB. In other words, by
combining two EB\ channels into a joint suitably-correlated environment, we
can reactivate the distribution of entanglement. We will show the physical
conditions under which the environmental correlations are able to trigger the
reactivation, therefore \textquotedblleft breaking
entanglement-breaking\textquotedblright. The most remarkable finding is that
we do not need to consider an entangled state for the environment: The
injection of separable correlations from the environment is sufficient for the restoration.

To better clarify these points, consider the schemes of direct and indirect
distribution in the presence of a correlated-noise environment. In the scheme
of direct distribution shown in panel~(3) of Fig.~\ref{scenario}, an input
entangled state $\rho_{AB}$ is jointly transformed into an output state
$\rho_{A^{\prime}B^{\prime}}=\mathcal{E}_{AB}(\rho_{AB})$. We assume that the
dilation of the composite channel $\mathcal{E}_{AB}$ is realized by
introducing a two-system environment, $E_{1}$ and $E_{2}$, in a bipartite
state $\rho_{E_{1}E_{2}}$, which interacts with the incoming systems via two
unitaries $U_{AE_{1}}$ (transforming $A$ and $E_{1}$) and $U_{BE_{2}}$
(transforming $B$ and $E_{2}$). In other words, the output state can be
written in the form
\begin{align}
\rho_{A^{\prime}B^{\prime}}  &  =\mathrm{Tr}_{E_{1}E_{2}}[(U_{AE_{1}}\otimes
U_{BE_{2}})\nonumber\\
&  \times(\rho_{AB}\otimes\rho_{E_{1}E_{2}})(U_{AE_{1}}^{\dagger}\otimes
U_{BE_{2}}^{\dagger})]. \label{eqDIL}%
\end{align}

If the environmental state is not tensor product, i.e., $\rho_{E_{1}E_{2}}%
\neq\rho_{E_{1}}\otimes\rho_{E_{2}}$, then the composite channel cannot be
decomposed into memoryless channels, i.e., $\mathcal{E}_{AB}\neq
\mathcal{E}_{A}\otimes\mathcal{E}_{B}$. In any case, from the dilation given
in Eq.~(\ref{eqDIL}), we can always define the reduced channels,
$\mathcal{E}_{A}$ and $\mathcal{E}_{B}$, acting on the individual systems. For
instance, if only system $B$ is transmitted, then we have the evolved state
\begin{align}
\rho_{AB^{\prime}}  &  =(\mathcal{I}_{A}\otimes\mathcal{E}_{B})(\rho
_{AB})\nonumber\\
&  =\mathrm{Tr}_{E_{2}}[(I_{A}\otimes U_{BE_{2}})\rho_{AB}\otimes\rho_{E_{2}%
}(I_{A}\otimes U_{BE_{2}}^{\dagger})],
\end{align}
where $\rho_{E_{2}}=\mathrm{Tr}_{E_{1}}(\rho_{E_{1}E_{2}})$. A similar formula
holds for the evolution of the other system $\rho_{A^{\prime}B}=(\mathcal{E}%
_{A}\otimes\mathcal{I}_{B})(\rho_{AB})$.

Now, assuming that $\mathcal{I}_{A}\otimes\mathcal{E}_{B}$ and $\mathcal{E}%
_{A}\otimes\mathcal{I}_{B}$ are EB channels (so that $A|B^{\prime}$ and
$B|A^{\prime}$), the composite channel $\mathcal{E}_{AB}$ can still preserve
entanglement (so that $A^{\prime}-B^{\prime}$ is possible). In other words, we
have a paradoxical situation where Charlie is not able to share entanglement
with Alice or Bob, but still can distribute entanglement to them. This is
clearly an effect of the injected correlations coming from the environmental
state $\rho_{E_{1}E_{2}}$. As mentioned earlier, our main finding is that
these correlations do not need to be strong: Entanglement distribution can be
activated by separable correlations, i.e., by an environment which is in a
separable state $\rho_{E_{1}E_{2}}$.


This effect of reactivation can also be extended to entanglement distillation,
which typically requires stronger conditions than entanglement distribution
(demanded by the existence of effective distillation protocols). Despite the
individual channels are EB, their combination into a separable environment
enables Charlie to distribute distillable entanglement to Alice and Bob. This
is easy to prove for an environment with finite memory, which can be
decomposed as $\mathcal{E}_{AB}\otimes\mathcal{E}_{AB}\otimes\ldots$

In our investigation, we also consider the case of entanglement swapping in a
correlated-noise environment as depicted in panel~(4) of Fig.~\ref{scenario}.
Here Alice and Bob have two entangled states, $\rho_{aA}$\ and $\rho_{Bb}$,
respectively. Systems $a$ and $b$ are retained, while systems $A$ and $B$ are
transmitted to Charlie, therefore undergoing the joint quantum channel
$\mathcal{E}_{AB}$. Before the Bell measurement, the global state is described
by
\begin{align}
\rho_{aA^{\prime}B^{\prime}b}  &  =(\mathcal{I}_{a}\otimes\mathcal{E}%
_{AB}\otimes\mathcal{I}_{b})(\rho_{aA}\otimes\rho_{Bb})\nonumber\\
&  =\mathrm{Tr}_{E_{1}E_{2}}[U(\rho_{aA}\otimes\rho_{E_{1}E_{2}}\otimes
\rho_{Bb})U^{\dagger}]~,
\end{align}
where $U=I_{a}\otimes U_{AE_{1}}\otimes U_{E_{2}B}\otimes I_{b}$.

As before, we consider the case where the reduced channels, $\mathcal{E}_{A}$
and $\mathcal{E}_{B}$, are EB channels, so that no entanglement survives
before the Bell measurement ($a|A^{\prime}$ and $B^{\prime}|b$). If the
environment has no memory ($\rho_{E_{1}E_{2}}=\rho_{E_{1}}\otimes\rho_{E_{2}}%
$) there is no way to distribute entanglement to Alice and Bob ($a|b$). By
contrast, if the environment has memory ($\rho_{E_{1}E_{2}}\neq\rho_{E_{1}%
}\otimes\rho_{E_{2}}$), then entanglement distribution is possible ($a-b$) and
this distribution can be activated by a separable environmental state
$\rho_{E_{1}E_{2}}$. Thus, we have the paradoxical situation where no
bipartite entanglement survives at Charlie's station ($a|A^{\prime}$ and
$B^{\prime}|b$), but still the swapping protocol is able to generate remote
entanglement at Alice's and Bob's stations ($a-b$) thanks to the separable
correlations injected by the environment. As before, these separable
correlations can be strong enough to distribute distillable entanglement to
the remote parties.

The environmental reactivation of entanglement distribution can be
proven~\cite{NJP} for quantum systems with Hilbert spaces of any dimension,
both finite (discrete-variable systems) and infinite (continuous-variable
systems~\cite{BraREV,BraREV2,RMP}). We remark that the phenomenon of
reactivation in direct distribution is not surprising in specific lossless
scenarios where the environment is \textquotedblleft
twirling\textquotedblright, i.e., a classical mixture of operators of the type
$U\otimes U$ or $U\otimes U^{\ast}$, with $U$ being a unitary~\cite{NJP}. In
this case, it is easy to find a fixed point in the joint map of the
environment, so that a state can be perfectly distributed, despite the fact
that the local (single-system) channels may become entanglement
breaking~\cite{NJP}. In discrete variables with Hilbert space dimensionality
$d\geq2$, these fixed points are the multi-dimensional Werner
states~\cite{Werner} (invariant under $U\otimes U$-twirling) and the
multi-dimensional isotropic states~\cite{HOROs} (invariant under $U\otimes
U^{\ast}$-twirling). Similarly, one can consider continuous-variable Werner
states which are invariant under anti-correlated phase-space rotations
(non-Gaussian twirlings)~\cite{NJP}. However, all these cases are artificial
since they are associated with lossless environments. The phenomenon of
reactivation becomes non-trivial in the presence of loss as typical for
continuous variable systems in realistic Gaussian environments.

In the configuration of indirect distribution, we can also find simple
examples of reactivation with discrete variable systems (in particular,
qubits) when the environment is lossless and $U\otimes U^{\ast}$-twirling.
Suppose that $\rho_{aA}$ and $\rho_{bB}$ are Bell pairs, e.g., singlet states
\begin{equation}
\left\vert -\right\rangle =\frac{1}{\sqrt{2}}\left(  \left\vert
0,1\right\rangle -\left\vert 1,0\right\rangle \right)  ~.
\end{equation}
Then suppose that qubits $A$ and $B$ are subject to twirling, which means that
$\rho_{AB}$ is transformed as
\begin{equation}
\rho_{A^{\prime}B^{\prime}}=\int dU~(U\otimes U^{\ast})~\rho_{AB}~(U\otimes
U^{\ast})^{\dagger}~,
\end{equation}
where the integral is over the entire unitary group $\mathcal{U}(2)$ acting on
the bi-dimensional Hilbert space and $dU$ is the Haar measure. Now the
application of a Bell detection on the output qubits $A^{\prime}$ and
$B^{\prime}$ has the effect to cancel the environmental noise. In fact, one
can easily check that the output state of $a$ and $b$ will be projected onto a
singlet state up to a Pauli operator, which is compensated via the
communication of the Bell outcome. Again, the phenomenon becomes non-trivial
when more realistic environments are taken into account, in particular, lossy
environments as typical for continuous-variable systems.

For this reason we discuss here the reactivation phenomenon using
continuous-variable systems. In particular, we consider the bosonic modes of
the electromagnetic field. The input modes are prepared in Gaussian states
with Einstein-Podolsky-Rosen (EPR) correlations~\cite{RMP,EPR}, which are the
most typical form of continuous variable entanglement. These modes are then
assumed to evolve under the action of a lossy Gaussian environment. This type
of environment is modelled by two beam splitters which mix the travelling
modes, $A$ and $B$, with two environmental modes, $E_{1}$ and $E_{2}$,
prepared in a bipartite Gaussian state $\rho_{E_{1}E_{2}}$ (separable or
entangled). The reduced channels, $\mathcal{E}_{A}$ and $\mathcal{E}_{B}%
$,\ are two lossy channels whose transmissivities and thermal noises are such
to make them EB channels. To achieve simple analytical results, in this
manuscript we only consider the limit of large entanglement for the input states.

The paper is structured as follows. In Sec.~\ref{SECgauss} we characterize the
basic model of correlated Gaussian environment, which directly generalizes the
standard model of thermal-loss environment. We identify the physical
conditions under which the correlated Gaussian environment is separable or
entangled. In Sec.~\ref{SECdirect}, we study the direct distribution of
entanglement in the presence of the correlated Gaussian environment and
assuming the condition of one-system EB. We provide the regimes of parameters
under which remote entanglement is activated by the environmental correlations
(in particular, separable correlations) and the stronger regimes where the
generated remote entanglement is also distillable. This part is a review of
results already known in the literature~\cite{NJP}. Then, in
Sec.~\ref{SECindirect}, we generalize the theory of entanglement swapping to
the correlated Gaussian environment. We consider swapping and distillation of
entanglement, finding the regimes of parameters where these tasks are
successful despite the EB condition. Finally, Sec.~\ref{SECconclusion} is for
conclusion and discussion.

\section{Correlated Gaussian environment\label{SECgauss}}

We consider two beam splitters (with transmissivity $\tau$) which combine
modes $A$ and $B$ with two environmental modes, $E_{1}$ and $E_{2}$,
respectively. These ancillary modes are in a zero-mean Gaussian state
$\rho_{E_{1}E_{2}}$ symmetric under $E_{1}$-$E_{2}$ permutation. In the
memoryless model, the environmental state is tensor-product $\rho_{E_{1}E_{2}%
}=\rho\otimes\rho$, meaning that $E_{1}$ and $E_{2}$ are fully independent. In
particular, $\rho$ is a thermal state with covariance matrix (CM)
$\omega\mathbf{I}$, where the noise variance $\omega=2\bar{n}+1$ quantifies
the mean number of thermal photons $\bar{n}$\ entering the beam splitter. Each
interaction is then equivalent to a lossy channel with transmissivity $\tau$
and thermal noise $\omega$.

\begin{figure}[ptbh]
\vspace{-2.8cm}
\par
\begin{center}
\includegraphics[width=0.65\textwidth] {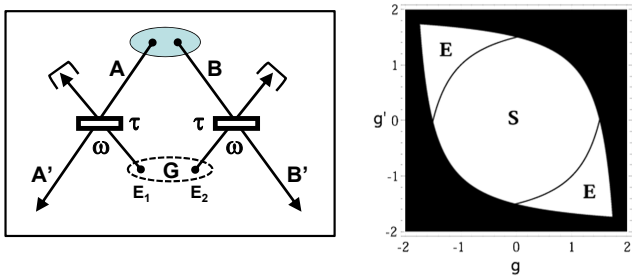}
\end{center}
\par
\vspace{-2.3cm}\caption{\textit{Left}. Correlated Gaussian
environment, with losses $\tau$, thermal noise $\omega$ and
correlations $\mathbf{G}$. The state of the environment $E_{1}$
and $E_{2}$ can be separable or entangled. \textit{Right}.
Correlation plane $(g,g^{\prime})$ for the Gaussian environment,
corresponding to thermal noise $\omega=2$. The black area
identifies forbidden environments (correlations are too strong to
be compatible with quantum mechanics). White area identifies
physical environments, i.e., the subset of points which satisfy
the bona-fide conditions of Eq.~(\ref{CMconstraints}). Whitin this
area, the inner region labbeled by S identifies separable
environments, while the two outer regions identify entangled
environments.  Figures adapted from Ref.~\cite{NJP} under a CC BY
3.0 licence (http://creativecommons.org/licenses/by/3.0/).}
\label{ENVschemes}
\end{figure}

This Gaussian process can be generalized to include the presence of
correlations between the environmental modes as depicted in the right panel of
Fig.~\ref{ENVschemes}. The simplest extension of the model consists\ of taking
the ancillary modes, $E_{1}$ and $E_{2}$, in a zero-mean Gaussian state
$\rho_{E_{1}E_{2}}$ with CM given by the symmetric normal form
\begin{equation}
\mathbf{V}_{E_{1}E_{2}}(\omega,g,g^{\prime})=\left(
\begin{array}
[c]{cc}%
\omega\mathbf{I} & \mathbf{G}\\
\mathbf{G} & \omega\mathbf{I}%
\end{array}
\right)  ~, \label{EVE_cmAPP}%
\end{equation}
where $\omega\geq1$ is the thermal noise variance associated with each
ancilla, and the off-diagonal block%
\begin{equation}
\mathbf{G=}\left(
\begin{array}
[c]{cc}%
g & \\
& g^{\prime}%
\end{array}
\right)  ~, \label{Gblock}%
\end{equation}
accounts for the correlations between the ancillas. This type of environment
can be separable or entangled (conditions for separability will be given afterwards).

It is clear that, when we consider the two interactions $A-E_{1}$ and
$B-E_{2}$ separately, the environmental correlations are washed away. In fact,
by tracing out $E_{2}$, we are left with mode $E_{1}$ in a thermal state
($\mathbf{V}_{E_{1}}=\omega\mathbf{I}$) which is combined with mode $A$ via
the beam-splitter. In other words, we have again a lossy channel with
transmissivity $\tau$ and thermal noise $\omega$. The scenario is identical
for the other mode $B$ when we trace out $E_{1}$. However, when we consider
the joint action of the two environmental modes, the correlation block
$\mathbf{G}$ comes into play and the global dynamics of the two travelling
modes becomes completely different from the standard memoryless scenario.

Before studying the system dynamics and the corresponding evolution of
entanglement, we need to characterize the correlation block $\mathbf{G}$\ more
precisely. In fact, the two correlation parameters, $g$ and $g^{\prime}$,
cannot be completely arbitrary but must satisfy specific physical constraints.
These parameters must vary within ranges which make the CM of
Eq.~(\ref{EVE_cmAPP}) a bona-fide quantum CM. Given an arbitrary value of the
thermal noise $\omega\geq1$, the correlation parameters must satisfy the
following three bona-fide conditions~\cite{TwomodePRA,NJP}
\begin{equation}
|g|<\omega,~~~|g^{\prime}|<\omega,~~~\omega^{2}+gg^{\prime}-1\geq
\omega\left\vert g+g^{\prime}\right\vert . \label{CMconstraints}%
\end{equation}

\subsection{Separability properties}

Once we have clarified the bona-fide conditions for the environment, the next
step is to characterize its separability properties. For this aim, we compute
the smallest partially-transposed symplectic (PTS) eigenvalue $\varepsilon$
associated with the CM\ $\mathbf{V}_{E_{1}E_{2}}$. For Gaussian states, this
eigenvalue represents an entanglement monotone which is equivalent to the
log-negativity~\cite{logNEG1,logNEG2,logNEG3} $\mathcal{E}=\max\left\{
0,-\log\varepsilon\right\}  $. After simple algebra, we get~\cite{NJP}%
\begin{equation}
\varepsilon=\sqrt{\omega^{2}-gg^{\prime}-\omega|g-g^{\prime}|}~.
\end{equation}
Provided that the conditions of Eq.~(\ref{CMconstraints}) are satisfied, the
separability condition $\varepsilon\geq1$ is equivalent to%
\begin{equation}
\omega^{2}-gg^{\prime}-1\geq\omega|g-g^{\prime}|~. \label{sepCON}%
\end{equation}

To visualize the structure of the environment, we provide a numerical example
in Fig.~\ref{ENVschemes}. In the right panel of this figure, we consider the
\textit{correlation plane} which is spanned by the two parameters $g$ and
$g^{\prime}$. For a given value of the thermal noise $\omega$, we identify the
subset of points which satisfy the bona-fide conditions of
Eq.~(\ref{CMconstraints}). This subset corresponds to the white area in the
figure. Within this area, we then characterize the regions which correspond to
separable environments (area labelled by S) and entangled environments (areas
labelled by E).

\section{Direct distribution of entanglement in a correlated Gaussian
environment\label{SECdirect}}

\begin{figure}[ptbh]
\vspace{-0.2cm}
\par
\begin{center}
\includegraphics[width=0.60\textwidth] {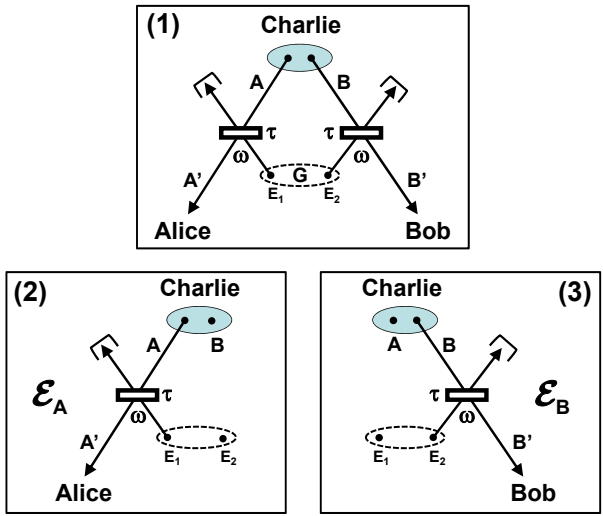}
\end{center}
\par
\vspace{-0.9cm}\caption{Scenarios for direct distribution of entanglement. (1)
Charlie has two modes $A$ and $B$ prepared in an EPR state $\rho_{AB}$. In
order to distribute entanglement to the remote parties, Charlie transmits the
two modes through the correlated Gaussian environment characterized by
transmissivity $\tau$, thermal noise $\omega$ and correlations $\mathbf{G}$.
(2) Charlie aims to share entanglement with Alice. He then keeps mode $B$
while sending mode $A$ to Alice through the lossy channel $\mathcal{E}_{A}$.
(3) \ Charlie aims to share entanglement with Bob. He then keeps mode $A$
while sending mode $B$ to Bob through the lossy channel $\mathcal{E}_{B}$.}%
\label{twomodeEB}%
\end{figure}

Let us study the system dynamics and the entanglement propagation in the
presence of a correlated Gaussian environment, reviewing some key results from
the literature~\cite{NJP}. Suppose that Charlie has an entanglement source
described by an EPR\ state $\rho_{AB}$ with CM%
\begin{equation}
\mathbf{V}(\mu)=\left(
\begin{array}
[c]{cc}%
\mu\mathbf{I} & \mu^{\prime}\mathbf{Z}\\
\mu^{\prime}\mathbf{Z} & \mu\mathbf{I}%
\end{array}
\right)  ~, \label{CM_TMSV}%
\end{equation}
where $\mu\geq1$, $\mu^{\prime}:=\sqrt{\mu^{2}-1}$, and $\mathbf{Z}$\ is the
reflection matrix%
\begin{equation}
\mathbf{Z}:=\left(
\begin{array}
[c]{cc}%
1 & \\
& -1
\end{array}
\right)  ~. \label{ZetaMAT}%
\end{equation}
We may consider the different scenarios depicted in the three panels of
Fig.~\ref{twomodeEB}. Charlie may attempt to distribute entanglement to Alice
and Bob as shown in Fig.~\ref{twomodeEB}(1), or he may try to share
entanglement with one of the remote parties, as shown in Figs.~\ref{twomodeEB}%
(2) and~(3).

Let us start considering the scenario where Charlie aims to share entanglement
with one of the remote parties (one-mode transmission). In particular, suppose
that Charlie wants to share entanglement with Bob (by symmetry the derivation
is the same if we consider Alice). For sharing entanglement, Charlie keeps
mode $A$ while sending mode $B$ to Bob as shown in Fig.~\ref{twomodeEB}(3).
The action of the environment is therefore reduced to $\mathcal{I}_{A}%
\otimes\mathcal{E}_{B}$, where $\mathcal{E}_{B}$ is a lossy channel applied to
mode $B$. It is easy to check~\cite{NJP} that the output state $\rho
_{AB^{\prime}}$, shared by Charlie and Bob, is Gaussian with zero mean and CM%
\begin{equation}
\mathbf{V}_{AB^{\prime}}=\left(
\begin{array}
[c]{cc}%
\mu\mathbf{I} & \mu^{\prime}\sqrt{\tau}\mathbf{Z}\\
\mu^{\prime}\sqrt{\tau}\mathbf{Z} & x\mathbf{I}%
\end{array}
\right)  , \label{ABpCM}%
\end{equation}
where
\begin{equation}
x:=\tau\mu+(1-\tau)\omega~.
\end{equation}

Remarkably, we can compute closed analytical formulas in the limit of large
$\mu$, i.e., large input entanglement. In this case, the entanglement of the
output state $\rho_{AB^{\prime}}$ is quantified by the PTS\ eigenvalue%
\begin{equation}
\varepsilon=\frac{1-\tau}{1+\tau}\omega~.
\end{equation}
The EB condition corresponds to the separability condition $\varepsilon\geq1$,
which provides%
\begin{equation}
\omega\geq\frac{1+\tau}{1-\tau}:=\omega_{\text{EB}}~, \label{EBcond}%
\end{equation}
or equivalently $\bar{n}\geq\tau/(1-\tau)$. Despite the EB condition of
Eq.~(\ref{EBcond}) regards an EPR\ input, it is valid for any input state. In
other words, a lossy channel $\mathcal{E}_{B}$\ with transmissivity $\tau$ and
thermal noise $\omega\geq\omega_{\text{EB}}$ destroys the entanglement of any
input state $\rho_{AB}$. Indeed Eq.~(\ref{EBcond}) corresponds exactly to the
well-known EB condition for lossy channels~\cite{HolevoEB}. The threshold
condition $\omega=\omega_{\text{EB}}$ guarantees one-mode EB, i.e., the
impossibility for Charlie to share entanglement with the remote party.

Now the central question is the following: Suppose that Charlie cannot share
any entanglement with the remote parties (one-mode EB), can Charlie still
distribute entanglement to them? In other words, suppose that the correlated
Gaussian environment has transmissivity $\tau$ and thermal noise
$\omega=\omega_{\text{EB}}$, so that the lossy channels $\mathcal{E}_{A}$\ and
$\mathcal{E}_{B}$ are EB. Is it still possible to use the joint channel
$\mathcal{E}_{AB}$ to distribute entanglement to Alice and Bob? In the
following, we explicitly reply to this question, discussing how entanglement
can be distributed by a separable environment, with the distributed amount
being large enough to be distilled by one-way distillation
protocols~\cite{NJP}.

Let us study the general evolution of the two modes $A$ and $B$ under the
action of the environment as in Fig.~\ref{twomodeEB}(1). Since the input
EPR\ state $\rho_{AB}$ is Gaussian and the environmental state $\rho
_{E_{1}E_{2}}$ is Gaussian, the output state $\rho_{A^{\prime}B^{\prime}}$ is
also Gaussian. This state has zero mean and CM given by~\cite{NJP}%
\begin{equation}
\mathbf{V}_{A^{\prime}B^{\prime}}=\tau\mathbf{V}_{AB}+(1-\tau)\mathbf{V}%
_{E_{1}E_{2}}=\left(
\begin{array}
[c]{cc}%
x\mathbf{I} & \mathbf{H}\\
\mathbf{H} & x\mathbf{I}%
\end{array}
\right)  ~,
\end{equation}
where%
\begin{equation}
\mathbf{H}:=\tau\mu^{\prime}\mathbf{Z}+(1-\tau)\mathbf{G}~.
\end{equation}
For large $\mu$, one can easily derive the symplectic spectrum of the output
state%
\begin{equation}
\nu_{\pm}=\sqrt{\left(  2\omega+g^{\prime}-g\pm|g+g^{\prime}|\right)
(1-\tau)\tau\mu}~,
\end{equation}
and its smallest PTS\ eigenvalue~\cite{NJP}%
\begin{equation}
\varepsilon=(1-\tau)\sqrt{(\omega-g)(\omega+g^{\prime})}~, \label{epsMAIN}%
\end{equation}
quantifying the entanglement distributed to Alice and Bob.

In the same limit, one can compute the coherent
information~\cite{CohINFO,CohINFO2} $I(A\rangle B)$ between the two remote
parties, which provides a lower bound to the number of entanglement bits per
copy that can be distilled using one-way distillation protocols, i.e.,
protocols based on local operations and one-way classical communication. It is
clear that one-way distillability implies two-way distillability, where both
forward and backward communication is employed. After simple algebra, one
achieves~\cite{NJP}%
\begin{equation}
I(A\rangle B)=\log\frac{1}{e\varepsilon}~. \label{coheDIR}%
\end{equation}
Thus, remote entanglement is distributed for $\varepsilon<1$ and is
distillable for $\varepsilon<e^{-1}$.

\begin{figure}[h]
\vspace{-0.0cm}
\par
\begin{center}
\includegraphics[width=0.48\textwidth] {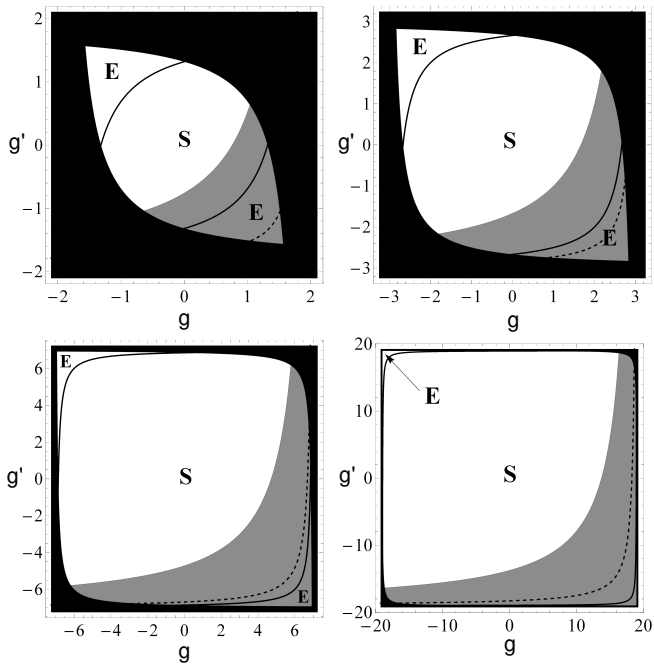}
\end{center}
\par
\vspace{-0.4cm}\caption{Analysis of the remote entanglement
$\varepsilon$ on the correlation plane $(g,g^{\prime})$ for
different values of the transmissivity $\tau=0.3$, $0.5$, $0.75$,
and $0.9$ (from top left to bottom right). Corresponding values of
the thermal noise are determined by the one-mode EB condition
$\omega=\omega_{\text{EB}}$. In each inset, the non-black area
identifies the set of physical environments, which are divided
into separable (S) and entangled (E) environments by the solid
lines. The gray region is the \textquotedblleft activation
area\textquotedblright\ and identifies those environments for
which Charlie is able to distribute entanglement to Alice and Bob
($\varepsilon<1$). Within the activation area, the environments
below the dashed curve are those for which the distributed
entanglement is also distillable by one-way protocols
($\varepsilon<e^{-1}$). Figure is re-used from Ref.~\cite{NJP}
under a CC BY 3.0 licence
(http://creativecommons.org/licenses/by/3.0/).}
\label{dirTOT}%
\end{figure}

Now suppose that the environment has thermal noise $\omega=\omega_{\text{EB}}$
(one-mode EB). Then, we can write
\begin{align}
\varepsilon &  =\sqrt{[1+\tau-(1-\tau)g][1+\tau+(1-\tau)g^{\prime}%
]}\nonumber\\
&  :=\varepsilon(\tau,g,g^{\prime}) \label{EBentEXP}%
\end{align}
Answering the previous question corresponds to checking the existence of
environmental parameters $\tau$, $g$ and $g^{\prime}$, for which\ $\varepsilon
$ is sufficiently low: For a given value of the transmissivity $\tau$, we look
for regions in the correlation plane $(g,g^{\prime})$ where $\varepsilon<1$
(remote entanglement is distributed) and possibly $\varepsilon<e^{-1}$ (remote
entanglement is distillable). This is done in Fig.~\ref{dirTOT} for several
numerical values of the transmissivity.

In Fig.~\ref{dirTOT}, the environments identified by the gray activation area
allow Charlie to distribute entanglement to Alice and Bob ($\varepsilon<1$),
despite it is impossible for him to share entanglement with any of the remote
parties. In other words, these environments are two-mode entanglement
preserving (EP), despite being one-mode EB. Furthermore, one can identify
sufficiently-correlated environments for which the entanglement distributed to
the remote parties can also be distilled ($\varepsilon<e^{-1}$).

The most remarkable feature in Fig.~\ref{dirTOT} is represented by the
presence of separable environments in the activation area. In other words,
there are separable environments which contain enough correlations to restore
the distribution of entanglement to Alice and Bob. Furthermore, for
sufficiently high transmissivities and correlations, these environments enable
Charlie to distribute distillable entanglement. As we can note from
Fig.~\ref{dirTOT}, the weight of separable environments in the activation area
increases for increasing transmissivities, with the entangled environments
almost disappearing for $\tau=0.9$.

\section{Entanglement swapping in a correlated Gaussian
environment\label{SECindirect}}

In this section we consider the indirect distribution of entanglement, i.e.,
the protocol of entanglement swapping. We start with a brief review of this
protocol in the ideal case of no noise. Then, we generalize its theory to the
case of correlated-noise Gaussian environments, where we prove how
entanglement swapping can be reactivated in the presence of one-mode EB.

\subsection{Entanglement swapping in the absence of noise\label{SECSUBswap1}}

\begin{figure}[ptbh]
\vspace{-1.6cm}
\par
\begin{center}
\includegraphics[width=0.6\textwidth] {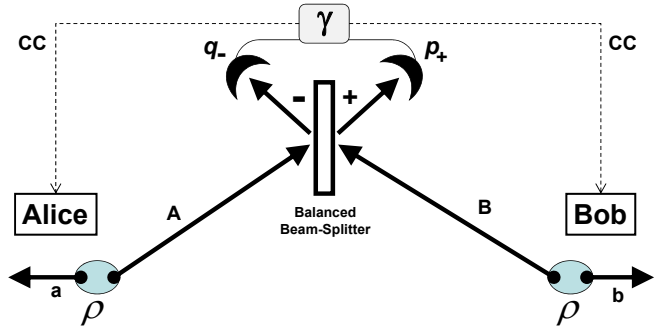}
\end{center}
\par
\vspace{-2.1cm}\caption{Entanglement swapping in the absence of noise. See
text for explanation.}%
\label{swap}%
\end{figure}

Consider two remote parties, Alice and Bob, who possess two identical EPR
states with CM given in Eq.~(\ref{CM_TMSV}). At Alice's station, the
EPR\ state describes modes $a$ and $A$, while at Bob's station it describes
modes $b$ and $B$. Alice and Bob keep modes $a$ and $b$, while sending modes
$A$ and $B$ to Charlie, where a Bell measurement is performed. This means that
the travelling modes $A$ and $B$ are combined in a balanced beam splitter
whose output modes \textquotedblleft$-$\textquotedblright\ and
\textquotedblleft$+$\textquotedblright\ are homodyned, with mode
\textquotedblleft$-$\textquotedblright\ measured in the position quadrature
and mode \textquotedblleft$+$\textquotedblright\ in the momentum quadrature.
In other words, Charlie measures the two EPR quadratures $\hat{q}_{-}%
:=(\hat{q}_{A}-\hat{q}_{B})/\sqrt{2}$ and $\hat{p}_{+}:=(\hat{p}_{A}+\hat
{p}_{B})/\sqrt{2}$. The Bell measurement provides two classical outcomes,
$q_{-}$ and $p_{+}$, which can be compacted into a single complex variable
$\gamma:=q_{-}+ip_{+}$. The classical variable $\gamma$ is finally
communicated to Alice and Bob, with the result of projecting their remote
modes $a$ and $b$ into a conditional state $\rho_{ab|\gamma}$ (see
Fig.~\ref{swap}). 

Since the input states are pure Gaussian and the Bell measurement is a
Gaussian measurement which projects pure states into pure states, we have that
the remote conditional state $\rho_{ab|\gamma}$ turns out to be a pure
Gaussian state. This state has a measurement-dependent mean $\mathbf{x}%
=\mathbf{x}(\gamma)$ which Alice and Bob can always delete by conditional
displacements. It is clear that these local unitaries do not alter the amount
of entanglement in the state, as long as they are perfectly implemented. The
conditional CM $\mathbf{V}_{ab|\gamma}$ can be computed using a simple
input-output formula for Gaussian entanglement swapping~\cite{GaussSWAP}. We
get
\begin{equation}
\mathbf{V}_{ab|\gamma}=\frac{1}{2\mu}\left(
\begin{array}
[c]{cc}%
(\mu^{2}+1)\mathbf{I} & (\mu^{2}-1)\mathbf{Z}\\
(\mu^{2}-1)\mathbf{Z} & (\mu^{2}+1)\mathbf{I}%
\end{array}
\right)  ~. \label{CMswapNLess}%
\end{equation}
Its smallest PTS\ eigenvalue is equal to $\varepsilon=\mu^{-1}$, which means
that remote entanglement is always generated for entangled inputs ($\mu>1$).
Furthermore, remote entanglement is present in the form of EPR correlations
since the two remote EPR quadratures $\hat{q}_{-}^{r}:=(\hat{q}_{a}-\hat
{q}_{b})/\sqrt{2}$ and $\hat{p}_{+}^{r}:=(\hat{p}_{a}+\hat{p}_{b})/\sqrt{2}$
have variances%
\begin{equation}
V(\hat{q}_{-}^{r})=V(\hat{p}_{+}^{r})=\mu^{-1}~.
\end{equation}

The simplest description of the entanglement swapping protocol can be given
when we consider the limit for $\mu\rightarrow\infty$. In this case the
initial states are ideal EPR states with quadratures perfectly correlated,
i.e., $\hat{q}_{a}=\hat{q}_{A}$ and $\hat{p}_{a}=-\hat{p}_{A}$ for Alice, and
$\hat{q}_{b}=\hat{q}_{B}$ and $\hat{p}_{b}=-\hat{p}_{B}$ for Bob. Then, the
overall action of Charlie, i.e., the Bell measurement plus classical
communication, corresponds to create a remote state with
\begin{equation}
\hat{q}_{b}=\hat{q}_{a}-\sqrt{2}q_{-},~\hat{p}_{b}=-\hat{p}_{a}-\sqrt{2}%
p_{+}~.
\end{equation}
The quadratures of the two remote modes are perfectly correlated, up to an
erasable displacement. In other words, the ideal EPR\ correlations have been
swapped from the initial states to the final conditional state $\rho
_{ab|\gamma}$.

\subsection{Entanglement swapping in the presence of
correlated-noise\label{SECSUBswap2}}

\begin{figure}[ptbh]
\vspace{-1.5cm}
\par
\begin{center}
\includegraphics[width=0.6\textwidth] {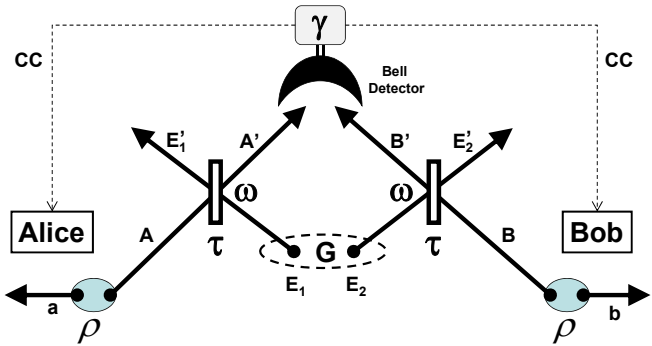}
\end{center}
\par
\vspace{-1.9cm}\caption{Entanglement swapping in the presence of loss, thermal
noise and environmental correlations (correlated Gaussian environment). The
Bell detector has been simplified.}%
\label{swapLOSS}%
\end{figure}

The theory of entanglement swapping can be extended to include the presence of
loss and correlated noise. We consider our model of correlated Gaussian
environment with transmission $\tau$, thermal noise $\omega$ and correlations
$\mathbf{G}$. The modified scenario is depicted in Fig.~\ref{swapLOSS}%
.

\subsubsection{Swapping of EPR\ correlations}

For simplicity, we start by studying the evolution of the EPR correlations
under ideal input conditions ($\mu\rightarrow+\infty$). After the classical
communication of the outcome $\gamma$, the quadratures of the remote modes $a$
and $b$ satisfy the asymptotic relations
\begin{align}
\hat{q}_{b}  &  =\hat{q}_{a}-\sqrt{\frac{2}{\tau}}\left(  q_{-}-\sqrt{1-\tau
}\hat{\delta}_{q}\right)  ,\label{eqrrr}\\
\hat{p}_{b}  &  =-\hat{p}_{a}-\sqrt{\frac{2}{\tau}}\left(  p_{+}-\sqrt{1-\tau
}\hat{\delta}_{p}\right)  , \label{eqrrr2}%
\end{align}
where $\hat{\delta}_{q}=(\hat{q}_{E_{1}}-\hat{q}_{E_{2}})/\sqrt{2}$ and
$\hat{\delta}_{p}=(\hat{p}_{E_{1}}+\hat{p}_{E_{2}})/\sqrt{2}$ are noise
variables introduced by the environment.

Using previous Eqs.~(\ref{eqrrr}) and~(\ref{eqrrr2}), we construct the remote
EPR\ quadratures $\hat{q}_{-}^{r}$\ and $\hat{p}_{+}^{r}$, and we compute the
EPR variances%
\begin{equation}
\boldsymbol{\Lambda}:=\left(
\begin{array}
[c]{cc}%
V(\hat{q}_{-}^{r}) & \\
& V(\hat{p}_{+}^{r})
\end{array}
\right)  \rightarrow\boldsymbol{\Lambda}_{\infty}=\frac{1-\tau}{\tau}%
(\omega\mathbf{I}-\mathbf{ZG})~, \label{lam1}%
\end{equation}
where the limit is taken for $\mu\rightarrow+\infty$. Assuming the EB
condition $\omega=\omega_{\text{EB}}$, we finally get%
\begin{equation}
\boldsymbol{\Lambda}_{\infty,\text{EB}}=\frac{1}{\tau}\left[  (1+\tau
)\mathbf{I}-(1-\tau)\mathbf{ZG}\right]  . \label{lam2}%
\end{equation}

In the case of a memoryless environment ($\mathbf{G=0}$) we see that
$\boldsymbol{\Lambda}_{\infty,\text{EB}}=(1+\tau^{-1})\mathbf{I}\geq
\mathbf{I}$, which means that the EPR\ correlations cannot be swapped to the
remote systems. However, it is evident from Eq.~(\ref{lam2}) that there are
choices for the correlation block $\mathbf{G}$\ such that the EPR\ condition
$\boldsymbol{\Lambda}_{\infty,\text{EB}}<\mathbf{I}$ is satisfied. For
instance, this happens when we consider $\mathbf{G=}g\mathbf{Z}$. In this case
it is easy to check that $\boldsymbol{\Lambda}_{\infty,\text{EB}}<\mathbf{I}$
is satisfied for $\tau\geq1/4$ and $g>(1-\tau)^{-1}$. Under these conditions,
EPR\ correlations are successfully swapped to the remote modes. In particular,
for $\tau>1/2$ and $g>(1-\tau)^{-1}$ there are separable environments which do
the job.

\subsubsection{Swapping and distillation of entanglement}

Here we discuss in detail how entanglement is distributed by the swapping
protocol in the presence of a correlated Gaussian environment. In particular,
suppose that Alice and Bob cannot share entanglement with Charlie because the
environment is one-mode EB. Then, we aim to address the following questions:
(i)~Is it still possible for Charlie to distribute entanglement to the remote
parties thanks to the environmental correlations? (ii)~In particular, is the
swapping successful when the environmental correlations are separable?
(iii)~Finally, are Alice and Bob able to distill the swapped entanglement by
means of one-way distillation protocols? Our previous discussion on EPR
correlations clearly suggests that these questions have positive answers. Here
we explicitly show this is indeed true for quantum entanglement by finding the
typical regimes of parameters that the Gaussian environment must satisfy.

In order to study the propagation of entanglement we first need to derive the
CM\ $\mathbf{V}_{ab|\gamma}$\ of the conditional remote state $\rho
_{ab|\gamma}$. As before, we have two identical EPR\ states at Alice's and
Bob's stations with CM $\mathbf{V}(\mu)$ given in Eq.~(\ref{CM_TMSV}). The
travelling modes $A$ and $B$ are sent to Charlie through a Gaussian
environment with transmissivity $\tau$, thermal noise $\omega$ and
correlations $\mathbf{G}$. After the Bell measurement and the classical
communication of the result $\gamma$, the conditional remote state at Alice's
and Bob's stations is Gaussian with CM~\cite{BellFORMULA}%
\begin{equation}
\mathbf{V}_{ab|\gamma}=\left(
\begin{array}
[c]{cc}%
\mu\mathbf{I} & \\
& \mu\mathbf{I}%
\end{array}
\right)  -\frac{(\mu^{2}-1)\tau}{2}\left(
\begin{array}
[c]{cccc}%
\frac{1}{\theta} &  & -\frac{1}{\theta} & \\
& \frac{1}{\theta^{\prime}} &  & \frac{1}{\theta^{\prime}}\\
-\frac{1}{\theta} &  & \frac{1}{\theta} & \\
& \frac{1}{\theta^{\prime}} &  & \frac{1}{\theta^{\prime}}%
\end{array}
\right)  ~, \label{VabGamma}%
\end{equation}
where%
\begin{equation}
\theta=\tau\mu+(1-\tau)(\omega-g),~\theta^{\prime}=\tau\mu+(1-\tau
)(\omega+g^{\prime})~. \label{thetas}%
\end{equation}

From the CM of Eq.~(\ref{VabGamma}) we compute the smallest PTS eigenvalue
$\varepsilon$ quantifying the remote entanglement at Alice's and Bob's
stations. For large input entanglement $\mu\gg1$, we find a closed formula in
terms of the environmental parameters, i.e.,%
\begin{equation}
\varepsilon=\frac{1-\tau}{\tau}\sqrt{(\omega-g)(\omega+g^{\prime}%
)}:=\varepsilon(\tau,\omega,g,g^{\prime})~, \label{SpectrumTOT}%
\end{equation}
which is equal to Eq.~(\ref{epsMAIN}) up to a factor $\tau^{-1}$. As before,
this eigenvalue not only determines the log-negativity but also the coherent
information $I(a\rangle b)$\ associated with the remote state $\rho
_{ab|\gamma}$. In fact, for large $\mu$, one can easily compute the asymptotic
expression%
\begin{equation}
I(a\rangle b)\rightarrow\log\frac{2}{e}\sqrt{\frac{\det\mathbf{V}_{b|\gamma}%
}{\det\mathbf{V}_{ab|\gamma}}}=\log\frac{1}{e\varepsilon}~,
\label{IabCOHEswap}%
\end{equation}
which is identical to the formula of Eq.~(\ref{coheDIR}) for the case of
direct distribution. Thus, the PTS\ eigenvalue of Eq.~(\ref{SpectrumTOT})
contains all the information about the distribution and distillation of
entanglement in the swapping scenario. For $\varepsilon<1$ entanglement is
successfully distributed by the swapping protocol (log-negativity
$\mathcal{E}>0$). Then, for the stronger condition $\varepsilon<e^{-1}$, the
swapped entanglement can also be distilled into $I(a\rangle b)$ entanglement
bits per copy by means of one-way protocols.

\begin{figure}[ptbh]
\vspace{-0.1cm}
\par
\begin{center}
\includegraphics[width=0.49\textwidth] {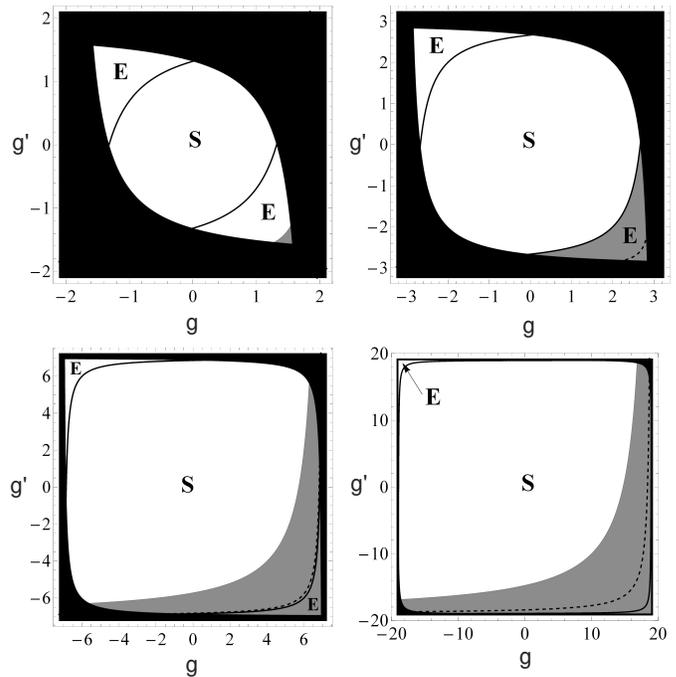}
\end{center}
\par
\vspace{-0.5cm}\caption{Analysis of the swapped entanglement $\varepsilon$ in
the correlation plane $(g,g^{\prime})$ for different values of the
transmissivity $\tau=0.3$, $0.5$, $0.75$, and $0.9$ (from top left to bottom
right). Corresponding values of the thermal noise are determined by the
one-mode EB condition $\omega=\omega_{\text{EB}}$. In each inset, the
non-black area identifies the set of physical environments, which are divided
into separable (S) and entangled (E) environments. The gray region identifies
the activation area, i.e., those environments whose correlations are strong
enough to generate remote entanglement ($\varepsilon<1$). Within the gray
activation area, the environments that lie below the dashed curve are those
able to generate distillable entanglement ($\varepsilon<e^{-1}$).}%
\label{total}%
\end{figure}

Now, let us assume the condition of one-mode EB ($\omega=\omega_{\text{EB}}$)
so that the bipartite states before measurement $\rho_{aA^{\prime}}$ and
$\rho_{B^{\prime}b}$ are separable (see Fig.~\ref{swapLOSS}). We investigate
the amount of entanglement generated in the remote modes $a$ and $b$ by
computing the eigenvalue $\varepsilon(\tau,\omega_{\text{EB}},g,g^{\prime})$.
In the standard memoryless case ($\mathbf{G}=\mathbf{0}$) we have
$\varepsilon=1+\tau^{-1}$ which means that no entanglement can be swapped, as
expected. To study the general case of correlated environment, we consider
different numerical values of the transmissivity $\tau$, and we plot the
$\varepsilon(\tau,\omega_{\text{EB}},g,g^{\prime})$ on the correlation plane.
The results are shown in Fig.~\ref{total} and are similar to those achieved in
Fig.~\ref{dirTOT} for direct distribution.

In each panel of Fig.~\ref{total}, the physical values for the correlation
parameters $(g,g^{\prime})$ are individuated by the non-black area. Remote
entanglement is distributed ($\varepsilon<1$) for values of the correlation
parameters belonging to the gray activation area. For $\tau\leq1/2$ (top two
panels), we see that the activation area is confined within the region of
entangled environments. The property that entangled environments are necessary
for the reactivation of entanglement swapping at any $\tau\leq1/2$ is easy to
prove. In fact, suppose that $\varepsilon<1$ holds. By using its formula in
Eq.~(\ref{SpectrumTOT}) and the bona-fide conditions on the correlation
parameters given in Eq.~(\ref{CMconstraints}), we can write $\varepsilon
^{2}<1$ as
\begin{equation}
\omega^{2}-gg^{\prime}+\omega(g^{\prime}-g)<\left(  \frac{\tau}{1-\tau
}\right)  ^{2}~. \label{eqkk}%
\end{equation}
Now, for $\tau\leq1/2$, we have $\tau^{2}(1-\tau)^{-2}\leq1$ and using this
inequality in Eq.~(\ref{eqkk}), we derive
\begin{equation}
\omega^{2}-gg^{\prime}-1<\omega(g-g^{\prime})\leq\omega|g-g^{\prime}|~,
\end{equation}
which is the entanglement condition for the environment [i.e., the violation
of Eq.~(\ref{sepCON})].

It is clear that the most interesting result holds for transmissivities
$\tau>1/2$. In this regime, in fact, the distribution of remote entanglement
can be activated by separable environments. As explicitly shown for
$\tau=0.75$ and $0.9$, the activation area progressively invades the region of
separable environments. In other words, separable correlations become more and
more important for increasing transmissivities. Furthermore, for $\tau
\gtrsim0.75$, separable environments are even able to activate the
distribution of distillable entanglement ($\varepsilon<e^{-1}$). By comparing
Fig.~\ref{dirTOT} and Fig.~\ref{total}, we see how entanglement is more easily
generated and distilled by the direct protocol. This is a consequence of the
extra factor $\tau^{-1}$ in Eq.~(\ref{SpectrumTOT}), whose influence becomes
less important only at high transmissivities ($\tau\simeq1$).

\section{Conclusion\label{SECconclusion}}

In conclusion, we have investigated the distribution of entanglement in the
presence of correlated-noise Gaussian environments, proving how the injection
of separable correlations can recover from entanglement breaking. In order to
derive simple analytical results we have considered here only the case of
large entanglement for the input states. We have analyzed scenarios of direct
distribution and indirect distribution, i.e., entanglement swapping.
Surprisingly, the injection of the weaker separable correlations is sufficient
to restore the entanglement distribution, as we have shown for wide regimes of
parameters. Furthermore, the generated entanglement can be sufficient to be
distilled by means of one-way protocols. The fact that separability can be
exploited to recover from entanglement breaking is clearly a paradoxical
behavior which poses fundamental questions on the intimate relations between
local and nonlocal correlations.

\section*{Acknowledgements}

This work was funded by a Leverhulme Trust research fellowship, and the EPSRC
via `qDATA' (Grant No. EP/L011298/1) and the UK Quantum Communications Hub
(Grant No. EP/M013472/1).

\end{document}